\documentclass[prd,showpacs,amsmath,amssymb,nofootinbib]{revtex4}

\usepackage[cp866]{inputenc}
\usepackage[russian,english]{babel}
\usepackage[T2A]{fontenc}
\usepackage{amsmath}
\usepackage{graphicx}
\newcommand{\Fst}{{\mathop {\rule{0pt}{0pt}{F}}\limits^*}\rule{0pt}{0pt}}

\begin{document}

\title{Self-interaction in the Bopp-Podolsky electrodynamics:\\
 Can the observable mass of a charged particle depend on its acceleration?}
\author{Alexei E. Zayats}
\email{Alexei.Zayats@kpfu.ru} \affiliation{Department of General
Relativity and Gravitation, Kazan Federal University, Kremlevskaya
Street 18, Kazan 420008, Russia}
%
%\preprint{\bf version 1.3}
%
%
%\date{\today}% It is always \today, today,
%             %  but any date may be explicitly specified
%
\begin{abstract}
In this paper we obtain the expression for the self-force in the
model with the Lagrangian containing additional terms, quadratic
in Maxwell tensor derivatives (so-called Bopp-Podolsky
electrodynamics). Features of this force are analyzed for various
limiting cases. When a charged particle moves along straight line
with a uniform acceleration, an explicit formula is found. In the
framework of the considered model, an observable renormalized
particle mass is shown to depend on its acceleration. This
dependence allows, in principle, to extract experimentally a value
of the particle bare mass.
\end{abstract}
% PACS, the Physics and Astronomy
%                             % Classification Scheme.
%\keywords{nonminimal coupling, traversable wormhole}%Use showkeys class option if keyword
%                              %display desired
\maketitle

\section{Introduction}

Effects of the quantum field theory in the low-energy limit can be
described by the action functional with an effective Lagrangian,
which contains additional nonlinear terms and terms with higher
derivatives. Furthermore, in recent years the dark energy problem
and the accelerated expansion of the Universe have inspired an
interest in the various phenomenological models in cosmology,
which use Lagrangians of such a type (see, e.g.,
\cite{phencosmmodels}). Therefore, a reasonable selection of the
effective Lagrangian and a search of constraints on its
constituent parameters become needed.

A general form of the effective Lagrangian describing the
interaction of gravitational and electromagnetic fields, which is
composed of invariants containing derivatives up to the fourth
order, can be given as
\begin{align}
{\cal L}_{\rm eff}&=\frac{R}{2\kappa}+\frac{1}{16\pi}F_{ik}F^{ik}+
c^{(1)}_{\rm MG}R^2+c^{(2)}_{\rm
MG}R_{ik}R^{ik}\nonumber\\
{}&+c^{(1)}_{\rm NM}RF_{ik}F^{ik}+c^{(2)}_{\rm
NM}R_i^kF_{km}F^{im}+ c^{(3)}_{\rm
NM}R_{ikmn}F^{ik}F^{mn}{}\nonumber\\{}&+c^{(1)}_{\rm
NL}(F_{ik}F^{ik})^2+c^{(2)}_{\rm NL}(F_{ik}\Fst^{ik})^2+{c}_{\rm
BP}\nabla_iF^{im}\,\nabla^kF_{km} \,.\label{Leff}
\end{align}
The first two terms in this expression form the Lagrangian of the
standard (or {\it minimal}\/) Einstein-Maxwell model. Next two
invariants (with phenomenological constants $c^{(1)}_{\rm MG}$ and
$c^{(2)}_{\rm MG}$) relate to the various modified theories of
gravity \cite{fourthorder1}. Next three cross-terms with coupling
parameters $c^{(1)}_{\rm NM}$, $c^{(2)}_{\rm NM}$, and
$c^{(3)}_{\rm NM}$ describe the nonminimal interaction of gravity
and the electromagnetic field (see \cite{We} for history,
references, and the latest results). The last terms with
parameters $c^{(1)}_{\rm NL}$, $c^{(2)}_{\rm NL}$ and $c_{\rm BP}$
are associated with the nonlinear \cite{BornInfeld,HeisEul} and
nonlocal self-interaction of the electromagnetic field,
respectively. Note that in Eq.(\ref{Leff}) we omit terms which can
be reduced to already indicated invariants by identical
transformations and/or dropping a total 4-divergence.

When we examine the effects relating to particle motion in {\it
weak} gravitational fields, one can neglect the terms containing
the curvature tensor. In this case, the Lagrangian of the theory
reduces to
\begin{align}
{\cal L}_{\rm eff}&=\frac{1}{16\pi}F_{ik}F^{ik}+c^{(1)}_{\rm
NL}(F_{ik}F^{ik})^2+c^{(2)}_{\rm NL}(F_{ik}\Fst^{ik})^2+{c}_{\rm
BP}\nabla_iF^{im}\,\nabla^kF_{km} \,.\label{Leff1}
\end{align}
Moreover, among the remaining terms we focus on the last one only,
while the invariants of the fourth order with respect to the
Maxwell tensor components leading to the nonlinearity of the
equations will not be considered here for simplicity.

The invariant $\nabla_iF^{im}\,\nabla^kF_{km}$ gives rise to
significant physical effects, such as a recoil force caused by the
own electromagnetic field of charged particles (i.e., self-force).
The electrodynamic model extended by using such terms was first
considered independently by Bopp and Podolsky
\cite{Bopp,Podolsky1}. Unlike the Maxwell electrodynamics, this
modification has two essential features: firstly, an electric
field of the point particle possesses a finite energy, secondly,
besides the ``usual'' massless photon, the Bopp-Podolsky (BP)
electrodynamics predicts its ``heavy'' analog, and, in this sense,
the BP-model has a close relationship with the Pauli-Villars
regularization procedure (see, e.g., \cite{Kvasnica}) and Lee-Wick
model \cite{LeeWick}. A ``heavy photon'' mass is the only
uncertain parameter of this theory.

The work is organized as follows. In Section~\ref{BPformalism} we
recall some facts relating to the BP-electrodynamics. In
Section~\ref{BPselfforce} the expression for the self-force in
BP-model is considered in detail: Subsection~\ref{sfA} covers its
general properties, while in Subsection~\ref{sfB} we focus on the
toy-model describing a uniformly accelerated particle. We
summarize the results of the paper in Section~\ref{Discussion}.

Throughout this paper we use the units $c = \hbar= 1$ and assume
that the spacetime is flat and possesses the metric $g_{ik}={\rm
diag}(1,-1,-1,-1)$. According to this assumption, all covariant
derivatives are replaced below with ordinary partial ones.

\section{Bopp-Podolsky electrodynamics}\label{BPformalism}

\subsection{General formalism}

The electromagnetic field in the generalized Bopp-Podolsky
electrodynamics is described by the Lagrangian
\cite{Bopp,Podolsky1}
\begin{equation}\label{action}
    {\cal
    L}_{BP}=\frac{1}{16\pi}F_{ik}F^{ik}-\frac{1}{8\pi\mu^2}\,\partial_iF^{im}\,\partial^kF_{km}+A_ij^i\,.
\end{equation}
Here $A_i$ is the electromagnetic field potential,
$F_{ik}=\partial_iA_k-\partial_kA_i$ is the Maxwell tensor, and
$j^i$ denotes the 4-vector of the current density. Note this model
returns to the ordinary Maxwell electrodynamics when $\mu$ tends
to infinity.

The variation of the action functional with the Lagrangian
(\ref{action}) with respect to the potentials $A_i$ gives the
electromagnetic field equation
\begin{gather}
\partial_m\left(F^{mi}+\mu^{-2}\,G^{mi}\right)=4\pi
j^i\,,\nonumber\\
G_{mn}\equiv
\partial_m\partial^kF_{kn}-\partial_n\partial^kF_{km}. \label{Meq}
\end{gather}
If we take into account the Lorenz gauge condition $\partial_m
A^m=0$, it reduces to
\begin{equation}\label{Meq1}
    \left(1+\mu^{-2}\square\right)\square A_n=4\pi j_n,\qquad
    \square\equiv \partial_m\partial^m.
\end{equation}
As was shown in \cite{Bopp}, any solution of (\ref{Meq1}) can be
represented as a difference of two auxiliary field potentials
$A_i'$ and $A_i''$:
\begin{equation}\label{bi}
    A_n=A_n'-A_n'',
\end{equation}
where the first term is a solution to the Maxwell equation, while
the second one obeys the Proca equation
\begin{equation}
\square A_{n}'=4\pi j_{\,n},\qquad \square A_{n}''+\mu^2A_n''=4\pi
j_{\,n}.\label{Meq2}
\end{equation}
Thus, the fourth-order equation (\ref{Meq1}) splits into two
second-order equations describing, correspondingly, massless and
massive vector fields with the same source. In this context, the
positive parameter $\mu$ in the BP-model plays the role of an
auxiliary field mass.

The energy-momentum tensor of the electromagnetic field in this
model takes the form \cite{Podolsky1}
\begin{equation}
T_{ik}^{BP}=\frac{1}{4\pi}\left(\frac14 F_{mn}F^{mn}
    g_{ik}-F_{im}{F_k}^m\right)+\mu^{-2}\,T_{ik}^{(+)}.\label{TEI0}
\end{equation}
Here the first component is the standard energy-momentum tensor in
the Maxwell theory and the additional term $T_{ik}^{(+)}$ is
defined as follows
\begin{gather}
    T_{ik}^{(+)}=\frac{1}{4\pi}\left(\frac{1}{2}\,g_{ik}\,\partial_pF^{pm}\,\partial^qF_{qm}
    -\partial^mF_{mi}\partial^nF_{nk}+\frac{1}{2}\,G_{mn}F^{mn}g_{ik}-G_{im}{F_k}^m-G_{km}{F_i}^m\right)\,.\label{Tplus}
\end{gather}
The substitution of (\ref{bi}) into (\ref{TEI0}) gives that the
energy-momentum tensor is also splitting into two parts
\cite{Bopp}:
\begin{equation}\label{TEI}
    T_{ik}=T_{ik}'-T_{ik}'',
\end{equation}
where
\begin{gather}
    T_{ik}'=\frac{1}{4\pi}\left(\frac14 F_{mn}'F'^{\,mn}
    g_{ik}-F_{im}'{F_k'}^m\right),\\
    F_{mn}'\equiv\partial_mA_n'-\partial_nA_m'\nonumber
\end{gather}
corresponds to the massless vector field $A_i'$, and
\begin{gather}
    T_{ik}''=\frac{1}{4\pi}\left(\frac14 F_{mn}''F''^{\,mn}
    g_{ik}-F_{im}''{F_k''}^m+\mu^2\left[A_i''A_k''-\frac{1}{2}g_{ik}A_m''A''^m\right]\right),\\
    F_{mn}''\equiv\partial_mA_n''-\partial_nA_m'',\nonumber
\end{gather}
is the energy-momentum tensor of the Proca-type massive field
$A_i''$. It is necessary to emphasize that this auxiliary field
has a negative energy and therefore the massive field $A_i''$ can
be interpreted as a {\it ghost} (or {\it phantom}) field.

\subsection{Point electric charge field}

\subsubsection{Charged particle at rest}

Let us consider the field of a point charged particle. When the
particle is at rest, its current density 4-vector takes the form
(hereafter we apply bold letters to denote spatial vectors)
\begin{equation}
    j^{\,0}=q\,\delta({\bf r}),\qquad {\bf j}=0,
\end{equation}
where $q$ is a charge of the particle, $\delta(x)$ is the Dirac
delta function. Solving Eqs.(\ref{Meq2}), we obtain the field
potential for the static case in the BP-electrodynamics
\cite{Bopp,Lande2,Podolsky1} as follows
\begin{equation}\label{statpoten}
    A_0=\frac{q}{r}(1-{\rm e}^{-\mu r})\,.
\end{equation}
For $r\gg 1/\mu$ the formula (\ref{statpoten}) turns into the
well-known expression for the Coulomb electrostatic potential,
$A_0=\frac{q}{r}$; while at the origin it remains nonsingular,
$\lim\limits_{r\to0}A_0=\mu q$.  Notice that the same formula for
the potential can be derived in the framework of the Maxwell
electrodynamics, if one supposes the electric charge has a spatial
distribution with density $j^{\,0}=\frac{\mu^2 q}{4\pi r}\,{\rm
e}^{-\mu r}$. In that case, the quantity $1/\mu$ plays the role of
a particle effective ``radius''. The energy $m_{\rm em}^{(0)}$ of
the electrostatic field (\ref{statpoten}) in the BP-model is also
finite and expressed by the formula
\begin{equation}
    m_{\rm em}^{(0)}=\int dV\, T_{00}=\frac{1}{2}\,\mu\,q^2.\label{E0}
\end{equation}
The obtained value can be identified as an electromagnetic
component of the mass of the rest charged particle, or, for
simplicity, its ``electromagnetic'' mass.

\subsubsection{Moving charged particle}

Let us proceed to a field configuration produced by a point charge
moving along a given trajectory ${\bf r}={\bf r}_0(t)$ with a
speed ${\bf v}_0(t)$. In this case, the components of the current
density 4-vector take the form
\begin{equation}\label{movcharge}
    j^{\,0}({\bf r},t)=q\, \delta({\bf r}-{\bf r}_0(t)),\quad {\bf j}({\bf r},t)=q\, {\bf v}_0(t) \delta({\bf r}-{\bf
    r}_0(t)).
\end{equation}
As was shown by Land\'e and Thomas \cite{Lande2}, the solution to
Eqs.(\ref{Meq2}) is given by the formula
\begin{equation}
 A_i=\mu
q\int\limits_0^{+\infty}d\xi\,{\rm J}_1(\mu\xi)\,
\frac1\xi\frac{\partial R_i}{\partial\xi},\label{Acharge}
\end{equation}
where $R^i$ denotes a position vector of the particle,
$R^i=x^i-x^i_{0}=(t-\tau,{\bf r}-{\bf r}_0(\tau))$, the
nonnegative parameter $\xi$ is defined by the relation
$\xi^2=R_pR^p$, $\tau$ is a retarded time which is determined
implicitly by the equation $\tau=t-\sqrt{\xi^2+|{\bf r}-{\bf
r}_0(\tau)|^2}$, and ${\rm J}_1(x)$ is the Bessel function of the
first order. Note that this expression represents an BP-model
analog of the Li\'{e}nard-Wiechert potentials in the Maxwell
electrodynamics.

Using standard methods and the identity $\partial_i
R_k=g_{ik}-R_i\,\frac{1}{\xi}\,\frac{\partial
    R_k}{\partial\xi}$,
we arrive to the expression for the strength tensor of the
electromagnetic field (\ref{Acharge})
\begin{equation}
    F_{ik}=\mu q\int\limits_0^{+\infty}d\xi\,{\rm J}_1(\mu\xi)\,\frac1\xi\frac{\partial}{\partial\xi}\left[R_k\,\frac1\xi\frac{\partial
R_i}{\partial\xi}-R_i\,\frac1\xi\frac{\partial
R_k}{\partial\xi}\right].\label{Fcharge}
\end{equation}
Thus, if the law of motion for the charged particle is known, its
electromagnetic field can be always derived by (\ref{Acharge}) and
(\ref{Fcharge}). In the next section, using  Eq.(\ref{Fcharge}),
we will obtain the formula for the self-force in
BP-electrodynamics.

\subsection{On the value of the parameter $\mu$}

There exist several values for the constant $\mu$, which have
already been proposed in the literature. On the one hand, the
Bopp-Podolsky additional term in the Lagrangian is an integral
part of the effective Lagrangian, which takes into account quantum
field corrections. The calculation of the electron-positron vacuum
polarization effect in the one-loop approximation of the quantum
electrodynamics yields \cite{Drum,Kvasnica}
\begin{equation}
    \mu=\sqrt{\frac{15\pi}{\alpha}}\,m_e\approx 80\,
    m_e.\label{muqed}
\end{equation}
Here $\alpha$ is the fine structure constant, $m_e$ is the
electron mass.

On the other hand, the BP-model can be considered in the
phenomenological context. Basing on the naive assumption that a
whole observable mass of the lightest charged particle (an
electron) consists only of the field component $m_{\rm em}^{(0)}$
only, we find
\begin{equation}
\mu=\frac{2}{\alpha}\,m_e\approx 274\,m_e\,.\label{mufull}
\end{equation}
In the general case, the above value can be considered just as the
{\it upper} bound for the parameter $\mu$.

Furthermore, in the literature there exist few attempts to define
the value of $\mu$, based on experimental data
\cite{Bonin,Cuzinatto2,Kvasnica2,Accioly}. However, all of these
estimations gives a lower limit and vary from $\mu\gtrsim
10^{-9}m_e$ till $\mu>70\,m_e$ (see \cite{Cuzinatto2}). Thus,
since there are no well-determined values of the parameter $\mu$,
we do not specify it through this paper.

\section{Self-force}\label{BPselfforce}

\subsection{General case}\label{sfA}

The self-force phenomenon for charged particles moving in the flat
spacetime was elaborated and described in detail (see, e.g.,
\cite{Rohrlich,selfaction} and references therein). The origin of
the particle self-force is related with the inertial properties of
the electromagnetic field. In other words, the self-force is the
reaction of radiation. The equation of motion for a charged
particle, which is under the influence of an external force
$F^i_{\rm ext}$, taking into account self-force effects yields
\begin{equation}\label{eqmotion}
    m\,\omega^i=F^i_{\rm ext}+f^i_{\rm sf},
\end{equation}
where $m$ is the particle mass, $\omega^i$ denotes 4-vector of its
acceleration, and $f^i_{\rm sf}$ is the self-force.

In contrast to the Maxwell electrodynamics, in the BP-model the
expression of the point charge field potential (\ref{statpoten})
is free from a singularity at the origin. Therefore, to derive a
formula for $f^i_{\rm sf}$, we can omit a renormalization
procedure and write down immediately
\begin{equation}
f^i_{\rm sf}=q F^{ik}u_k,\label{selfforcedef}
\end{equation}
where $F^{ik}$ is the Maxwell tensor of the electromagnetic field
produced by the moving particle, and $u_k$ is the 4-vector of its
velocity, and both quantities have to be calculated at the present
location of the charged particle. The substitution of the formula
for the strength tensor (\ref{Fcharge}) into (\ref{selfforcedef})
yields
\begin{align}
f^i_{\rm sf}&=\mu q^2\int\limits_0^{+\infty}d\xi\,{\rm
J}_1(\mu\xi)\,\frac1\xi\frac{\partial}{\partial\xi}\left[(R_ku^k)\,\frac1\xi\frac{\partial
R^i}{\partial\xi}-R^i\,\frac1\xi\frac{\partial
(R_ku^k)}{\partial\xi}\right].\label{selfforce}
\end{align}
The structure of this expression points to essential nonlocality
of the self-force, because the whole path traversed by the
particle up to the present time contributes to $f^i_{\rm sf}$
($\xi^2=R_pR^p$ runs from 0 to $+\infty$).

In order to analyze the general formula (\ref{selfforce}) for an
arbitrary type of motion, let us consider the power expansion of
$f^i_{\rm sf}$ in terms of the parameter $\mu$, using the
asymptotic formula from \cite{asympint}
\begin{align}
    \int\limits_0^{+\infty}\frac{{\rm J}_1(\mu
    x)}{x}y(x)\,dx=
    y(0)+\frac{1}{\mu}y'(0)+\frac{1}{2\mu^2}y''(0)-\frac{1}{8\mu^4}y^{(4)}(0)+\ldots.\label{asympexpan}
\end{align}
If $\xi=0$ we have either $\tau=t$, or $\frac{|{\bf r}_0(t)-{\bf
r}_0(\tau)|}{t-\tau}=1$. But the last relation is impossible,
because massive particles cannot move with the speed being more or
equal to the speed of light. Hence, $\lim\limits_{\xi\to 0}R^i=0$.

Using this fact, let us obtain the expansion of the position
vector $R^i$ with respect to $\xi$. The particle trajectory will
be considered to be parameterized by a natural parameter $s$. At
the present point $s=s_0$, while the retarded moment $\tau$
corresponds to $s=s_0-\Delta s$, where $\Delta s>0$. It is easy to
see that the expansion of $R^i$ in terms of $\Delta s$ takes the
form
\begin{align}
R^{\,i}=x^i(s_0)-x^i(s_0-\Delta s)=u^i \Delta
s{-}\frac{1}{2}\,\omega^i\,\Delta s^2{+}\frac{1}{6}\dot{\omega}^i
\Delta s^3{-}
    \frac{1}{24}\ddot{\omega}^i \Delta s^4{+}\ldots,\label{S}
\end{align}
where $\omega^i=du^i/ds$ is the particle acceleration, the dot
denotes the derivative with respect to $s$, and the quantities
$u^i$, $\omega^i$, etc are calculated at the present point of the
particle, i.e., at $s=s_0$. From the condition $R_iR^i=\xi^2$ we
find the relation, which connects the parameters $\Delta s$ and
$\xi$
\begin{equation}\label{S1}
\xi^2=\Delta s^2-\frac{1}{12}\,\omega_i\omega^i\,\Delta
s^4+\frac{1}{12}\omega_i\dot{\omega}^i\Delta s^5+\ldots\,.
\end{equation}
Solving this equation with respect to $\Delta s$ and taking into
account that $\Delta s>0$, we arrive at
\begin{equation}\label{S2}
    \Delta s=\xi+\frac{1}{24}\omega_i\omega^i\xi^3-\frac{1}{24}\omega_i\dot{\omega}^i\xi^4+\ldots\,.
\end{equation}
The substitution of this relation into (\ref{S}) yields
\begin{eqnarray}\label{S3}
    R^{\,i}=u^i\xi-\frac{1}{2}\,\omega^i\,\xi^2+\frac{1}{6}\left(\dot{\omega}^i+\frac{1}{4}u^i\omega^k\omega_k\right)\xi^3-
    \frac{1}{24}\left(\ddot{\omega}^i+u^i\dot{\omega}^k\omega_k+\omega^i\omega^k\omega_k\right)\xi^4+\ldots\,.
\end{eqnarray}
Hence, using (\ref{S3}) and (\ref{asympexpan}), we obtain at last
that
\begin{align}
    f^i_{\rm sf}=-\frac{\mu
    q^2}{2}\omega^i+\frac{2q^2}{3}\left(\dot\omega^i+u^i\omega^k\omega_k\right)
    -\frac{3q^2}{8\mu}\left(\ddot\omega^i+3u^i\dot\omega^k\omega_k+\frac{3}{2}\omega^i\omega^k\omega_k\right)+\dots\,.\label{selfexpan}
\end{align}
Note that this formula was found earlier by McManus
\cite{McManus}, who was guided by quite different motivation.

The expansion (\ref{selfexpan}) is just a formal one, because it
is unknown whether this series converges. Nevertheless, basing on
(\ref{selfexpan}), one can discover a number of specific
properties of the self-force $f^i_{\rm sf}$.

When $\mu\to\infty$, dropping the first infinite term, which
corresponds to the classical renormalization of the mass, the
expression (\ref{selfexpan}) transforms into the well-known
formula for the Lorentz-Dirac force
\begin{equation}\label{LDselfforce}
    f^i_{\rm LD}=\frac{2q^2}{3}\left(\dot\omega^i+u^i\omega^k\omega_k\right).
\end{equation}
But, in contrast to the Maxwell electrodynamics, the first term
$\frac{\mu q^2}{2}\,\omega^i$ in (\ref{selfexpan}) being
proportional to the acceleration of the moving particle is finite.
The factor is equal to the electromagnetic field energy of the
pointlike particle at rest (see (\ref{E0})), and this feature is
in agreement with Frenkel's work \cite{Frenkel1}, where this
problem was considered in the nonrelativistic approximation.

On the other hand, besides terms, which are known from the usual
Lorentz-Dirac theory, in the formula (\ref{selfexpan}) there exist
additional terms depending on the higher derivatives of the
acceleration $\omega^i$. They vanish in the limit $\mu\to\infty$,
but these terms can exert essential influence on the behavior of
the self-force at the finite values of $\mu$. For example, one can
believe that due to these corrections there are no
self-accelerated, or runaway solutions in the BP-model (see
\cite{Lande3,Frenkel2}).

\subsection{Self-interaction in the case of uniformly accelerated motion}\label{sfB}

For the sake of simplicity, let a charged particle move along a
straight line under the influence of a constant external force,
being collinear to particle's motion. Then the motion equation of
this particle takes the form \cite{Rohrlich}
\begin{gather}
    t(s)=\frac{1}{w}\sinh ws,\quad x(s)=\frac{1}{w}\cosh ws,\quad y(s)=z(s)=0.
\end{gather}
Here we assume that the trajectory lies along the axis $Ox$ and
the natural parameter $s$ runs from $-\infty$ to $+\infty$. The
constant $w$ is a magnitude of the particle acceleration,
$w^2=-\omega_i\omega^i$.

In this case, the velocity $u^i$, the acceleration $\omega^i$, and
the position vector $R^i=x^i(s)-x^i(s-\Delta s)$ are of the form
\begin{gather}
    u^i=\cosh ws\,\delta^i_0+\sinh ws\,\delta^i_1, \\
    \omega^i=w\left(\sinh ws\,\delta^i_0+\cosh ws\,\delta^i_1\right), \\
    R^{\,i}=\frac{1}{w}\left[(\sinh ws-\sinh w(s-\Delta s))\,\delta^i_0+(\cosh ws-\cosh
    w(s-\Delta s))\,\delta^i_1\right].
\end{gather}
From the condition $R_pR^p=\xi^2$ we obtain that the parameters
$\xi$ and $\Delta s$ are connected by the following relations
\begin{gather}
   \cosh w\/\Delta s=1+\frac{w^2\xi^2}{2},\\ \sinh w\/\Delta s=w\xi
   \sqrt{1+\frac{w^2\xi^2}{4}}.\label{xis}
\end{gather}
Hence we arrive to the expression for the vector $R^i$ expressed
in terms of $\xi$
\begin{equation}\label{Rxi}
     R^i=\xi \sqrt{1+\frac{w^2\xi^2}{4}}\,
   u^i-\frac{\omega^i\xi^2}{2}.
\end{equation}
Substituting now (\ref{Rxi}) into the general formula for the
self-force (\ref{selfforce}), we find that in the case of
uniformly accelerated motion the vector $f^i_{\rm sf}$ is aligned
along the particle acceleration $\omega^i$
\begin{equation}
    f^i_{\rm sf}=-m_{\rm em}(w)\,\omega^i,\label{unaccsefforce}
\end{equation}
where the factor $m_{\rm em}(w)$ is given by the relation
\begin{align}
    m_{\rm em}(w)=\frac{\mu q^2}{2}\int\limits_0^{+\infty} \frac{d\xi}{\xi}\frac{{\rm J}_1(\mu
    \xi)}{\left(1+\frac{w^2\xi^2}{4}\right)^{3/2}}=\frac{\mu^2
    q^2}{w}\, {\rm I}_1\!\!\left(\frac{\mu}{w}\right) {\rm K}_1\!\!\left(\frac{\mu}{w}\right).
\end{align}
\begin{figure}[t]
\includegraphics[width=8cm]{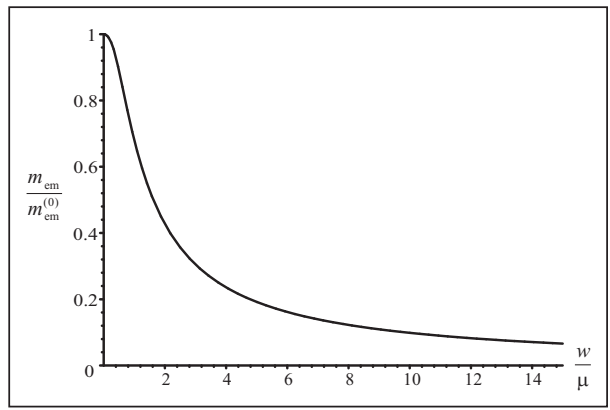}
\caption{Plot of the factor $m_{\rm em}$ normalized to the
electromagnetic mass of the rest particle against the
dimensionless parameter $w/\mu$. At the origin, i.e., when the
acceleration $w$ vanishes, $m_{\rm em}/m_{\rm em}^{(0)}=1$. The
function decreases monotonically and tends to zero at
infinity.}\label{figure}
\end{figure}
Here ${\rm I}_1(x)$ and ${\rm K}_1(x)$ are the modified Bessel
functions of the first and second kind, respectively. As it is
easy to see, the quantity $m_{\rm em}$, which can be identified as
a ``electromagnetic'' mass of a moving particle, depends on the
magnitude of its acceleration $w$. If $w\to0$, then obviously
$m_{\rm em}\to m_{\rm em}^{(0)}=\frac{\mu q^2}{2}$. Otherwise, the
function $m_{\rm em}(w)$ approaches to zero when $w\to\infty$. To
illustrate the behavior of this function in the mentioned cases,
we present the plot (see Fig.~\ref{figure}) and two asymptotic
expansions ($\gamma$ denotes Euler's constant)
\begin{equation}
\frac{m_{\rm em}}{m_{\rm em}^{(0)}}=
1-\frac{3}{8}\left(\frac{w}{\mu}\right)^2+\dots,\ \hbox{if}\
\frac{w}{\mu}\ll 1,
\end{equation}
\begin{equation}
\frac{m_{\rm em}}{m_{\rm em}^{(0)}}=
\frac{2\mu}{w}{+}\left[\frac{1}{2}\ln\frac{\mu}{2w}{+}\frac{\gamma}{2}{-}\frac{1}{8}\right]\left(\frac{\mu}{w}\right)^2{+}
\dots,\ \hbox{if}\ \frac{w}{\mu}\gg 1.
\end{equation}

As the result, the equation of charged particle motion under
influence of the constant external force $F_{\rm ext}^i$ and the
self-force (\ref{unaccsefforce}) is written as
\begin{equation}
    \left[m+m_{\rm em}(w)\right]\omega^i=F_{\rm ext}^i\,.
\end{equation}
It is easy to see that the quantity $m_{\rm obs}=m+m_{\rm em}(w)$
plays the role of an observable mass of the particle. If the
particle acceleration vanishes, then the observable mass $m_{\rm
obs}=m+\frac{\mu q^2}{2}$ consists as expected of the proper, or
``bare'', mass of the particle $m$ and the term associated with
the electrostatic field energy (\ref{E0}). However, when the
particle is in uniformly accelerated motion, its observable mass
decreases and tends to $m$, if $w\to\infty$. This phenomenon opens
up a possibility to reveal, in principle, the bare mass of charged
particles.

\section{Conclusion}\label{Discussion}

The Bopp-Podolsky electrodynamics, which is the simplest
high-order modification of the Maxwell theory, was proposed in the
1940s, but it has been remaining topical (see, e.g., recent papers
\cite{Bonin,Accioly,Cuzinatto2,Maceda}). We can point out three
reasons to explain this fact:
\par 1. This model is free of divergences, since the potential and
the electrostatic field energy are finite.
\par 2. In addition to the phenomenological way, the emergence of high-order terms
in the Lagrangian can be predicted in the framework of the
quantum field theory.
\par 3. This version of the theory of electromagnetism is elaborated
insufficiently and offers the possibilities of {\it new physics}.
\par In the present paper we focus on one of such new phenomena.
From the toy-model discussed above it follows that in the
Bopp-Podolsky electrodynamics the interaction between a charged
particle and its electromagnetic field gives rise to the
self-force directed against the acceleration of the particle. A
related phenomenon known as a virtual mass effect occurs in
hydrodynamics, when a solid moves with an acceleration through the
liquid column (see, e.g., \cite{fluid}), but for our case the
factor $m_{\rm em}$ depends on the particle acceleration.

It is important to note that the equations given in
Subsection~\ref{sfB} have been obtained for a uniform movement of
the particle provided $t\in(-\infty;+\infty)$. Nevertheless, the
effect described above takes place even for arbitrary motion. In
order to explain this statement we consider the expansion
(\ref{selfexpan}) of the self-force $f_{\rm sf}^i$. Since it
contains terms like $\omega^i \omega^k\omega_k$, one can conclude
that the factor in front of the acceleration 4-vector, $m_{\rm
em}$, will depend on the particle acceleration for the general
case. Moreover, we can say that the observable mass depends not
only on the magnitude of the acceleration as in the formula
(\ref{unaccsefforce}), but also on its direction and derivatives
$\dot\omega^i$, $\ddot\omega^i$, etc. Thus, for a more realistic
example, we can consider particle motion within a plate-parallel
capacitor, and obtain dependence between the accelerating force
and the particle acceleration by experiments. As for the model
considered in Subsection~\ref{sfB}, it represents the simplest
example, for which we can calculate this effect explicitly.

Dependence between the observable mass and the particle
acceleration gives us a hypothetical possibility to obtain
experimentally the model parameter $\mu$ as well as the bare mass
of the charged particle, which is customary believed to be
unobservable. Experimental testing of the obtained laws allows to
define or put restrictions on parameter values in the effective
Lagrangian and, as a consequence, validity of one or another of
the approaches to produce it.

It is worth recalling that in our investigation we ignored the
terms quartic in the Maxwell tensor. This is legitimate provided
the magnitude of the neglected terms is much less than the rest.
For a field produced by a rest charge at $r=0$ we have
\begin{gather}
F_{ik}F^{ik}\sim \mu^4q^2,\quad
\frac{1}{\mu^2}\,\partial^iF_{ik}\,\partial_mF^{mk}\sim
\mu^4q^2,\quad (F_{ik}F^{ik})^2\sim \mu^8q^4.
\end{gather}
Hence we obtain that the disregard for the terms indicated by
constants $c_{\rm NL}^{(1)}$ and $c_{\rm NL}^{(2)}$ in
(\ref{Leff}) is justified for small values of $\mu$, namely,
\begin{equation}
    \mu^4\ll \frac{1}{q^2\,c_{\rm NL}^{(1,2)}}\,.
\end{equation}
Otherwise, for big values of the parameter $\mu$ it is necessary
to take into account the quartic terms. This work will be done by
us in future papers.

\appendix

\acknowledgments

This work was partially supported by the Russian Foundation for
Basic Research (Grant No. 11-02-01162) and the Ministry of
education and science of Russian Federation (project
No.~14.B37.21.2035).

\end{document}